\documentclass[useAMS,usenatbib]{mn2e}

\usepackage{epsfig}
\usepackage{amssymb}
\usepackage{amsmath}
              
\def\araa{{\it Annu. Rev. Astron. Astrophys.} \,}

\def\apj{{\it ApJ \,}}
\def\apjs{{\it Ap. J. Supp.} \,}

\def\apjl{{\it Ap. J. Lett.} \,}

\def\prd{{\it Phys. Rev. D.} \,}
\def\mnras{{\it MNRAS} \,}
\def\aj{{\it Astron. J.} \,}

\def\nat{{\it Nature} \,}
\def\aap{{\it A\&A} \,}
\def\araa{{\it } \,}
\def\pasj{{\it Publ. Astron. Soc. Japan} \,}

\begin{document}
\title[Cluster contribution to CXRB]{Cluster contribution to the X-ray background as a cosmological probe}

\author[Doron Lemze et al.]{Doron Lemze$^{1}$\thanks{E-mail: doronl@wise.tau.ac.il}, 
Sharon Sadeh$^1$, \& Yoel Rephaeli$^{1,2}$\\
$^{1}$School of Physics and Astronomy, Tel Aviv University, Tel Aviv, 69978, Israel\\
$^{2}$Center for Astrophysics and Space Sciences, University of California, San Diego\\}

\pagerange{\pageref{firstpage}--\pageref{lastpage}} \pubyear{2008}
\maketitle

\begin{abstract}

Extensive measurements of the X-ray background (XRB) yield a reasonably 
reliable characterisation of its basic properties. Having resolved most 
of the cosmic XRB into discrete sources, the levels and spectral shapes 
of its main components can be used to probe both the source populations 
and also alternative cosmological and large-scale structure models. Recent 
observations of clusters seem to provide evidence that clusters formed 
earlier and are more abundant than predicted in the standard $\Lambda$CDM 
model. This motivates interest in alternative models that predict enhanced 
power on cluster scales. We calculate predicted levels and spectra of the 
superposed emission from groups and clusters of galaxies in $\Lambda$CDM 
and in two viable alternative non-Gaussian ($\chi^2$) and early dark energy 
models. The predicted levels of the contribution of clusters to the XRB in 
the non-Gaussian models exceed the measured level at low energies and 
levels of the residual XRB in the $2-8$ keV band; these particular models 
are essentially ruled out. Our work demonstrates the diagnostic value of 
the integrated X-ray emission from clusters, by considering also its 
dependences on different metallicities, gas and temperature profiles, 
Galactic absorption, merger scenarios, and on a non-thermal pressure 
component. We also show that the XRB can be used for a upper limit for
the concentration parameter value.

\end{abstract}


\section{Introduction}
\label{Introduction}

The origin of the cosmic X-ray background (CXRB) is of considerable 
interest due to possible ramifications on the properties of the various 
classes of sources that contribute to it, such as clusters and other 
extragalactic sources. By estimating their CXRB contributions and 
accounting for Galactic absorption we can compare with measurements of 
the observed X-ray background (XRB). Since the discovery of the XRB 
(Giacconi et al. 1962) significant effort has been devoted to spectrally 
mapping the emission at energies in the range $10^{-1} - 10^{2}$ keV. 
The radiation has a power-law form with a spectral index of 
$\Gamma=1.4$ above $1$ keV; the spectrum $E\cdot I(E)$, where 
$I(E)$ is the spectral intensity, peaks at $E\sim 30$ keV and 
decreases exponentially at higher energies. Substantial uncertainties 
in the measurement of $I(E)$ are due to cross-calibration uncertainties 
(e.g., De Luce \& Molendi 2004) and spatial variations reflecting the 
large-scale structure (Gilli et al. 2003; Yang et al. 2003).

It has been determined that the XRB is mostly the superposed emissions 
from discrete extragalactic sources, other than at the lowest energies, 
$\lesssim 0.25$ keV, where its dominant contribution is thermal 
emission from the local bubble (LB) and the Galactic halo. Estimated 
fractional contributions of resolved sources to the XRB are 70-80\% in 
the ROSAT $0.5-2$ keV (Hasinger et al. 1998) and HEAO-1 A2 $2-10$ keV 
(Mushotsky et al. 2000) bands. Other estimates of the fractional 
contributions of resolved sources are 60-80\% (Giacconi et al. 2001), 
70\% in the 2-8 keV band (Cowie et al. 2002). More recently, Worsley 
et al. (2005) determined XRB fractions of ~85\%, ~80\%, ~60\%, ~50\% 
for the 0.5-2, 2-10, 6-10, 8-10 keV bands, respectively. Higher fractions 
were deduced from deep Chandra fields (CDFs), $89.5_{+5.9}^{-5.7}\%$ 
and $86.9_{+6.6}^{-6.3}\%$ of the in the 0.5-2 and 2-8 keV bands, 
respectively (Bauer et al. 2004). These contributions include AGN, which 
dominate above $1$ keV (Hasinger 2004, Brandt \& Hasinger 2005 and 
references therein; Gilli, Comastri, \& Hasinger 2007), star forming 
galaxies, and Galactic stars, but do not include emission from high-mass 
groups and clusters of galaxies (which will be simply referred to as 
clusters); see also (Hickox \& Markevitch 2006,2007). 

A small contribution at very low energies may come from the warm hot 
intergalactic medium (WHIM). Most of the unresolved 1-2 keV emission can 
be accounted for by galaxies that are detected in deep HST observations, 
but are too faint to be detected as individual X-ray sources, as shown by 
a stacking analysis (Worsley et al. 2006). Analysis of the CDFs implies 
that the maximum contribution of clusters is estimated 
to be $\sim 17$ \% and $\sim 20$ \% in the 0.5-2 keV and 2-8 keV 
bands, respectively. Clusters are important probes of the primordial 
density fluctuation spectrum and the evolution of the large scale 
structure. Since conditions in clusters can be gauged by levels of 
X-ray emission, their estimated contribution to the CXRB 
can be used as a statistical diagnostic measure of the cluster 
population and its evolution. 

The contribution of clusters to the XRB was estimated in several works 
(e.g., Rosati et al. 1998; Gilli, Risaliti,\& Salvati 1999; Wu \& Xue 
2001) to be less than $\sim 10\%$ of the XRB spectral intensity at 
$\sim 1$ keV, and even lower at energies $>2$ keV. The integrated 
spectral intensity of clusters can now be calculated more accurately 
from our improved knowledge of intracluster (IC) gas properties, the 
relevant scaling relations, as well as more precisely determined 
values of the global cosmological parameters. Moreover, X-ray line 
emission, which contributes significantly to the spectrum at low 
energies, particularly in high-mass groups, should be added to the 
thermal Bremsstrahlung emission (as has already been shown by Phillips, 
Ostriker, \& Cen 2001). 

Additional motivation for revisiting the issue of cluster contribution 
to the CXRB is provided by the possibility of using it to constrain 
alternative cosmological and LSS (Large scale structure ) models, that 
have been proposed to 
remove apparent tension between the predictions of the standard model 
and recent observations of clusters. Possibly discrepant results
include the following: (a) The detection of proto-spheroidal galaxies 
at $z\simeq 2$ with a substantially higher comoving density than what 
is predicted by most semi-analytic models of galaxy formation 
(Magliocchetti et al. 2007). Even more problematic is the detection of 
structures with high velocity dispersions at redshifts $z=4.1$ (Miley 
et al. 2004) and $z=2.1$ (Kurk et al. 2004). (b) The 
`excess' CMB power at high multipoles ($l\approx3000$) measured by the 
CBI (Readhead et al. 2004), ACBAR (Kuo et al. 2004), and BIMA (Dawson et 
al. 2006) arrays. The measured power levels seem to be too high to be 
comfortably consistent with the predicted CMB primary anisotropy, and 
are also outside the range of predicted anisotropy induced by the S-Z 
effect (Sadeh, Rephaeli, \& Silk 2006, 2007). (c) High values of the 
concentration parameter, which were deduced for A1689 from two independent 
analyses: a joint lensing and X-ray analysis (Lemze et al. 2008a, 
hereafter L08a), and a dynamical analysis (Lemze et al. 2008b, hereafter 
L08b) of this cluster. In addition, high concentrations can explain the 
recently measured large Einstein radii in five high-mass clusters 
(Broadhurst \& Barkana 2008), obtained from the analysis of high-quality 
ACS and Subaru lensing measurements and Chandra data. These values prove 
to be high even after relaxing the recent formation time assumption 
usually adopted in theoretical estimations, by considering the 
probability distribution function of halo formation times (Sadeh \& 
Rephaeli 2008).

All the above apparently discrepant observational results can be naturally 
explained in the context of LSS models with primordial density 
fluctuation spectrum that had excess power on cluster scales, resulting 
in earlier formation and higher abundances of high-mass clusters than 
in the standard $\Lambda$CDM model. Several such viable models with 
the desired enhanced power on cluster scales are non-Gaussian $\chi^2$ 
models (e.g., Koyama, Soda \& Taruya 1999) and early dark energy 
models with time-varying coefficient of the quintessence equation of state 
(Doran et al. 2001; Caldwell et al. 2003). It has already been suggested 
by Mathis, Diego, \& Silk (2004) that scale-dependent non-Gaussianity (at 
a level consistent with observational constraints from WMAP measurements) 
may be probed through its detectable manifestations on clusters. The 
predicted S-Z power spectra and cluster number counts were determined by 
Sadeh, Rephaeli, \& Silk (2006, 2007).

Earlier formation, higher abundances, and more concentrated IC gas 
profiles will also be reflected in possibly discernible differences in 
their X-ray properties when compared with those predicted in the standard 
$\Lambda$CDM model. Therefore, in addition to the need to update and 
improve calculations of the contribution of clusters to the CXRB, further 
motivation for extending our previous work on alternative cosmological 
models is provided by the expectation that cluster X-ray properties could 
yield additional useful constraints on these models.

This paper is arranged as follows: In \textsection~\ref{Methodology}
we describe the methodology adopted for the calculation of the spectral 
contribution of clusters and high mass groups in three distinct 
cosmological models. We present our results in 
\textsection~\ref{Results}, where predicted XRB levels are 
made for alternative cosmological models, different levels of Galactic 
absorption, various gas profiles, cluster mergers, and in the case when 
the total gas pressure includes a non-thermal component. We conclude with 
a discussion in \textsection~\ref{Discussion}.

\section{Methodology}
\label{Methodology}

X-ray emission from clusters is calculated in $\Lambda$CDM,   
non-Gaussian (NG), and early dark energy (EDE) models. In the NG 
models the probability distribution function (PDF) of the primordial 
density fluctuation field has a $\chi^{2}_m$ form. 
The largest deviation from Gaussianity occurs when there is one such field 
($m=1$). We consider the $m=1$ and $m=2$ cases. A full account of the 
the non-Gaussian and early dark energy models that we have used are 
provided in Sadeh, Rephaeli, \& Silk (2006, 2007). Here we present a 
brief summary of the most important aspects which are of particular 
relevance to this work.

\subsection{$\Lambda$CDM and Non-Gaussian Models} 
\label{LambdaCDM}

The number density of clusters as a function of mass and redshift is 
calculated using the Press-Schechter (Press \& Schechter 1974, hereafter 
PS) mass function 
\begin{equation}
\frac{dn(M)}{dM}=\frac{\overline{\rho}_m}{M}\frac{d\sigma}{dM}F(\nu),
\end{equation} 
where 
\begin{eqnarray}
F(\nu)=\left\{ \begin{array}{ll}
N\sqrt{\frac{1}{2\pi}}e^{-\frac{\nu^2}{2}}\frac{\nu}{\sigma} & 
\\ \\
N\frac{(1+\sqrt{2/m}\nu)^{m/2-1}}{(2/m)^{(m-1)/2}\Gamma(m/2)}e^{-m/2(1+
\sqrt{m/2}\nu)}\frac{\nu}{\sigma}&
\end{array}
\right.\, .
\end{eqnarray}
The top and bottom expressions refer to the Gaussian and $\chi^2_m$ 
models, respectively, and $N$ is a normalization factor, introduced in 
order to arrange that all the matter in the universe will be incorporated 
into halos. 
 
For the Gaussian, NG (m=1), and NG (m=2) models, $N=2, 0.31, 0.37$, 
respectively. Additional quantities appearing in the PS mass function 
include $\nu \equiv \frac{\delta_{c}(z)}{\sigma(M)}$, which 
denotes the critical density for collapse in units of the mass variance. 
The latter quantity is defined as $\sigma^{2}(M)=
\frac{1}{(2\pi)^3}\int d^{3}k \widetilde W_{TH}^{2}(kR)
P(k)$, where $\widetilde{W}_{TH}(kR)=\frac{3}{(kR)^3}[sin(kR)
-kR cos(kR)]$ is the Fourier transform of the Top-Hat window 
function, and the relation between $M$ and $R$ is 
$R = \left( \frac{3M}{4\pi\rho_{bg}} \right)^{1/3}$. 
The power spectrum is taken as 
$P(k)=A_pk^nT^2(k)$, with $n=1$ and $n=-1.8$ for the Gaussian and NG 
models, respectively.

Adiabatic CDM and isocurvature transfer functions are used in 
the Gaussian and NG models:
 \begin{eqnarray}
T(k)=\left\{\begin{array}{ll}
\frac{ln(1+2.34q)}{2.34q}[1+3.89q+(16.1q)^2+(5.46q)^3\\
+(6.71q)^4]^{-0.25}  \\ \\
(5.6q)^2[1+\frac{(40q)2}{1+215q+(16q)^2(1+0.5q)^{-1}} \\+
(5.6q)^{8/5}]^{-5/4}
\end{array}
\right.
\end{eqnarray} 
where $q\equiv k/(\Omega_m\,h^2)Mpc^{-1}$ (Bardeen et al. 1986), 
and $A_p$ is the normalization set by $\sigma_8$:
\begin{equation}
A_p=\frac{\sigma_8^2}{\frac
{1}{2\pi^2}\int_{0}^{\infty}dk k^{n+2}T^2(k)\widetilde W_{TH}^{2}(8k)}. 
 \end{equation}
The cosmological parameters were taken from the WMAP 3-year data, 
with $\Omega_\Lambda=0.76$, $\Omega_m=0.24$, $h=0.73$, and 
$\sigma_8=0.74$ (Spergel et al.\ 2007). Due to the steep dependence of 
the cluster mass function on $\sigma_8$, we have accounted for the higher 
value, $\sim 0.82$, inferred from WMAP5 (Komatsu et al. 2008). This point
is elaborated on in \textsection~\ref{Discussion}. The gas profile is 
taken to be a $\beta$ model, i.e. $n_{gas}(r)=n_{gas}(0)(1+
(\frac{r}{r_c})^2)^{-1.5\beta}$, where the central density is 
normalized such that the total gas mass constitutes a fraction $0.1$ of 
the total cluster mass included within the virial radius, and $r_c$ 
denotes the core radius. To allow for a temperature structure, the 
profile is taken to have a polytopic form,
$T(r)=T(0)(1+(\frac{r}{r_c})^2)^{-1.5\beta(\gamma-1)}$, 
where $\gamma$ is the polytopic index, and the central temperature 
can be derived under the assumption of hydrostatic equilibrium (HE), 
that yields 
\begin{equation}
T(r=0)=\frac{\mu m_p G}{3\beta\gamma k_B}\frac{M_{vir}}{r_{vir}}\left[1+
\left(\frac{r_{vir}}{r_c}\right)^2\right]^{1.5 \beta(\gamma-1)+1}
\left(\frac{r_{vir}}{r_c}\right)^{-2}, 
\end{equation} 
where $G$, $m_p$, $\mu$ and $k_B$ are the gravitational constant, 
proton mass, mean molecular weight, and the Boltzmann constant, 
respectively.

We first consider isothermal gas, $\gamma=1$, and take 
$C_{gas}\equiv\frac{r_{vir}}{r_{c}}=10$, $\beta=\frac{2}{3}$. 
Note that $C_{gas}$ is the virial radius in units of the gas core radius, 
essentially the analogous concentration parameter, where the virial 
radius is in units of the NFW (Navarro, Frenk, \& White 1997) scale 
radius. We consider (in \textsection~\ref{gas profiles}) a range of 
values of $C_{gas}$ and $\gamma$ in order to assess the dependence on 
these parameters. With the gas density and temperature profiles 
specified, we can calculate the spectral luminosity of a cluster with 
mass $M$ and redshift $z$ 
\begin{equation}
L_{M,z}(E)=\int_{0}^{r_{vir}} \varepsilon_{E} dV, 
\end{equation}
where $\varepsilon_{E}$ is the specific (spectral) emissivity in 
$erg\; sec^{-1}\; cm^{-3}\; kev^{-1}$. We compute the emissivity 
with the MEKAL plasma code and compare with results obtained when the 
emissivity includes only Bremsstrahlung continuum 
(Fig.~\ref{spectrum_using_MEKAL}). The calculation of the Gaunt factor 
was carried out using the analytic fitting formula of Itoh et al. (2000). 

We consider all systems in the mass range $1.1\cdot 10^{13}-
7.3\cdot10^{15}$ h$_{0.7}^{-1}$ $M_{\odot}$, which roughly includes 
all halos larger than a cD galaxy. Increasing the upper halo mass limit 
does not affect the results, since the PS formalism predicts an extremely 
low abundance of such massive halos. Indeed, the most massive 
currently known cluster, A370, has a mass ($2.93^{+0.36}_{-0.32}
\cdot 10^{15}$ h$_{0.7}^{-1}$ M$_{\odot}$, Broadhurst et al. 
2008) which is less than half the assumed upper end of this range.
The specific flux from all clusters in this mass range is
\begin{equation}
F(E)=\int_{0}^{z_{max}} dz\int_{M_{min}}^{M_{max}} 
L_{M,z}(E)\frac{dn(M,z)}{dMdz}\frac{1}{4\pi D_{L}^2(z)} dM, 
\end{equation}
where $D_{L}$ is the luminosity 
distance; in our calculations we take $z_{max}=3$.

\subsection{EDE Model}
\label{EDE}

The evolutionary history in the EDE model is different from 
that in the $\Lambda$CDM universe, for which we used the Gaussian and NG 
models. The vacuum (Quintessence) energy density, $\rho_Q(z)$, is 
redshift dependent,
\begin{equation}
\rho_Q=\rho_{Q_0}(1+z)^{3[1+\overline{\omega}(z)]},
\end{equation} where $\overline{\omega}(z)$ is the equation of state 
coefficient ($-1$ for a cosmological constant). 
In our calculations we adopt the parametrisation of Wetterich (2004): 
\begin{equation}
\overline{\omega}(z)=\frac{\omega_0}{1+u\log(1+z)},
\end{equation} and $u\equiv\frac{-3\omega_0}{\log\frac{1-
\Omega_e}{\Omega_e}+\log\frac{1-\Omega_m}{\Omega_m}}$, where 
$\Omega_e$ is the early quintessence density and $\omega_0$ is 
$\omega(z=0)$. 
The specific EDE model assumed here is that investigated 
by Bartelmann, Doran, \& Wetterich (2006; hearafter BDW), for which 
$\Omega_e=8\cdot 10^{-4}$ and $\omega_0=-0.99$.
 
Several cosmological quantities which are affected by the presence of 
early dark energy require re-evaluation. These include the (comoving) 
radial and angular diameter distances, linear growth factor,  
the critical density for spherical collapse, and the overdensity at 
virialization, all of which are explicitly calculated by Sadeh, 
Rephaeli, \& Silk (2007).

\section{Results}
\label{Results}

\subsection{Cluster contribution to the CXRB in $\Lambda$CDM}
\label{cluster_CXRB_contribution}

\begin{figure} 
\centering
\epsfig{file=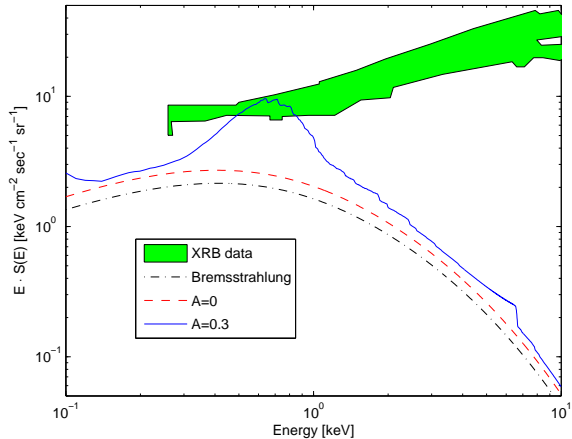, width=8.5cm, clip=}
\caption{The integrated contribution of clusters and high mass groups
with masses ($1.1\cdot10^{13}-7.3\cdot10^{15}$) h$_{0.7}^{-1}$ 
$M_{\odot}$ to the XRB intensity as a function of energy is shown 
together with the XRB data. Emission processes include free-free (black 
dash-dotted curve), optically thin hot plasma with zero metal abundance, 
A=0 (red dashed curve), and optically thin hot plasma with A=0.3 (blue 
solid curve). 
\label{spectrum_using_MEKAL}}
\end{figure}

As noted, we have used the MEKAL code to calculate gas emissivity. 
Inclusion of only Bremsstrahlung continuum emission appreciably 
underestimates radiation from hot plasmas, especially at the relatively 
low temperatures of the more abundant groups and poor clusters, for 
which line emission at low energies and continuum free-bound emission 
processes should also be included. 
In Fig.~\ref{spectrum_using_MEKAL} we 
show the superposed spectrum of clusters in the mass range 
$1.1\cdot10^{13}$ h$_{0.7}^{-1}$ $M_{\odot}$ - $7.3\cdot10^{15}$ 
h$_{0.7}^{-1}$ $M_{\odot}$ when including only free-free emissivity 
(dash-dotted line), optically thin hot plasma with zero metal abundances, 
A=0 (dashed line), and optically thin hot plasma with A=0.3 (solid line).
The XRB measurements (with uncertainties), shown here by the green 
region, are based on the compilation of Gilli et al. (2007), 
who combined measurements from HEAO-1 (Gruber 1992; Gruber et al. 1999), 
ASCA GIS (Kushino et al. 2002), ROSAT PSPC (Georgantopoulos et al. 1996), 
two different sets of XMM measurements 
(Lumb et al. 2002; De Luca \& Molendi 2004), ASCA SIS (Gendreau et al. 
1995), BeppoSAX (Vecchi et al. 1999), RXTE (Revnivtsev et al. 2003), and 
shadowing experiments (Warwick \& Roberts 1998). These measurements 
differ slightly in their minimum Galactic latitude. The effect of different minimum 
Galactic latitudes on the spectrum is investigated in 
\textsection~\ref{Galactic absorption}.
As can be inferred from Fig.~\ref{spectrum_using_MEKAL}, the level of 
the Bremsstrahlung emission alone is quite low in comparison with the 
XRB measurements. Its contribution peaks at $\sim 0.4$ keV, where it 
constitutes $\sim 20$\% of the CXRB. Inclusion of the full continuum 
(Bremsstrahlung and free-bound) with A=0 enhances the cluster emissivity 
by $\sim 1.25$ across the entire $0.1-10$ keV energy band. With the 
higher abundance of A=0.3, the emissivity is much higher, especially 
in the $0.4-1$ keV energy band.

\begin{figure} 
\centering
\epsfig{file=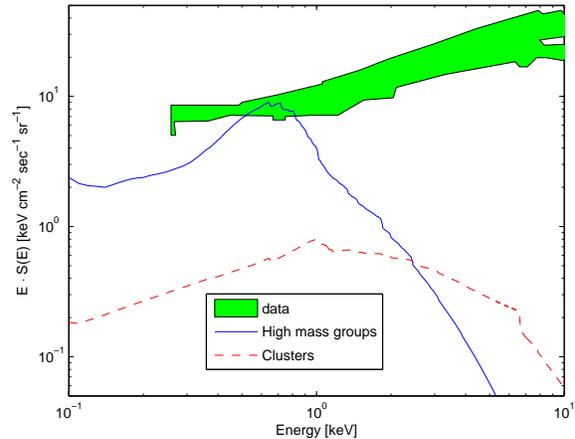, width=8.5cm, clip=}
\caption{The relative XRB spectral intensity contributed by high-mass 
groups and clusters, defined by the respective mass ranges of 
$1.1\cdot 10^{13}$ h$_{0.7}^{-1}$ M$_{\odot}$- $10^{14}$ h$_{0.7}^{-1}$ 
M$_{\odot}$ and $10^{14}$ h$_{0.7}^{-1}$ $M_{\odot}$ - $7.3\cdot10^{15}$ 
h$_{0.7}^{-1}$ M$_{\odot}$.
\label{highmassgroups_Vs_clusters}}
\end{figure}

Clearly, high mass groups and clusters contribute to the CXRB 
in different energy bands. Clusters contribute at higher energies 
due to their higher masses. We illustrate in Fig.~ 
\ref{highmassgroups_Vs_clusters} the contribution of the two populations, 
where we took the mass range of the high mass groups and clusters to be 
$1.1\cdot 10^{13}$ h$_{0.7}^{-1}$ M$_{\odot}$ - $10^{14}$ h$_{0.7}^{-1}$ 
M$_{\odot}$ and $10^{14}$ h$_{0.7}^{-1}$ $M_{\odot}$ - $7.3\cdot10^{15}$ 
h$_{0.7}^{-1}$ M$_{\odot}$, respectively. The former population 
dominates the emission below $\sim 2.5$ keV, whereas the latter 
contributes mostly at higher energies.
Clearly, the superposed emission of the lower mass systems can be used 
to constrain models that predict large spectral bumps, whereas the 
higher mass clusters can be used to constrain models in which fractional 
contribution of clusters is higher than the low residual 
level, namely upon subtraction of the dominant AGN contribution.

\begin{figure}
\centering
\epsfig{file=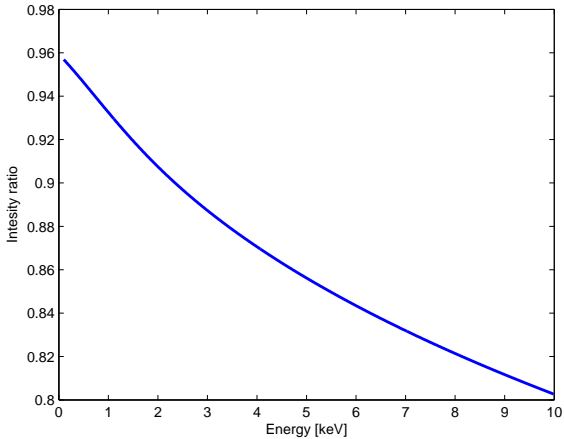, width=8.5cm, clip=}
\caption{The spectral intensity calculated using analytical fits for 
the linear growth factor, critical density for spherical collapse, and 
overdensity at virialization, compared with the corresponding spectral 
intensity calculated using the exact numerical solutions to the differential 
equations for these quantities. (This comparison was made for the case 
$A=0$.) 
\label{integrated_Vs_expression}}
\end{figure}

We have explored the dependence of the predicted intensity on the various 
LSS quantities and IC gas model. Here we consider the impact of using 
approximate fitting formulae for the linear growth factor, critical 
density for spherical collapse, and overdensity at virialization. To do 
so we numerically solved the relevant equations describing these 
quantities (see, e.g., Sadeh, Rephaeli, \& Silk 2007) and used these 
exact solutions instead of the respective approximate fitting formulae. 
We find that at low energies the two sets of calculations yield 
essentially the same results. However, with increasing energy the exact 
calculation yields a level of intensity which is lower by as much as 
$\sim 17\%$ at $10\,keV$,as can be seen in 
Fig.~\ref{integrated_Vs_expression}.

In the following subsections we assess the impact of IC gas modeling 
uncertainties, merger histories of clusters, and dependence of the 
predicted CXRB intensity on Galactic absorption.

\subsection{Galactic absorption} 
\label{Galactic absorption}

To account for photoelectric absorption in the Galaxy, 
we combined our calculated CXRB intensity with the XSPEC WABS model. 
We then computed the average column density in the following way: 
$10^{4}$ lines of sight (los) were picked at random with a uniform 
probability distribution on a sphere. This number proved to be sufficient 
in order to produce statistically meaningful results by virtue of the 
fact that with $10^3$ los the average column density changed negligibly 
and its error was still small. Note that the number of los has an upper 
limit of $41259$ since the resolution of the column density is $\sim 1^{\circ}$.

For each los we obtained the column density using the \emph{nh} tool 
(Dickey \& Lockman 1990). This procedure was repeated 10 times, yielding 
the average column density and its standard deviation, which was taken to 
be the error. In this way we calculated the average column density of the 
Galaxy in all directions. In order to take into account the fact that 
XRB observations are performed at different Galactic latitudes, we 
calculated the average Galactic absorption at stripes of Galactic 
latitude $b>10^{\circ}$ and $b>30^{\circ}$, namely excluding the 
disk region, and the combined regions of the disk and the bulge, 
respectively. In addition, many X-ray observations point at regions on 
the sky where the column density is significantly lower, the goal of 
which is the detection of the soft X-ray background (which is absorbed 
in regions of high column densities). Such a region is the Lockman Hole, 
associated with a minimum hydrogen column density 
$N_H=(4.5\pm0.5)\cdot 10^{19}\, cm^{-2}$ (Lockman et al. 1986).

\begin{figure} 
\centering
\epsfig{file=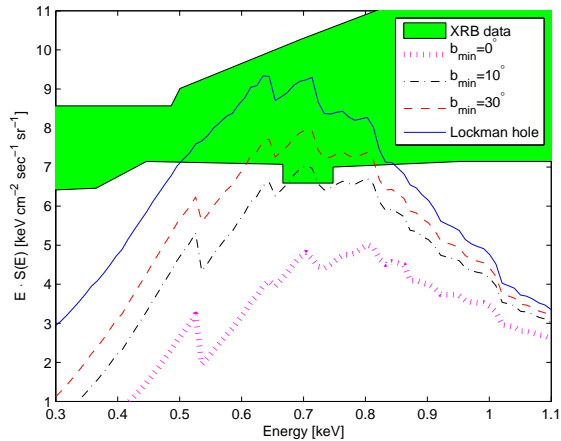, width=8.5cm, clip=}
\caption{The impact of Galactic absorption on the integrated contribution 
of clusters and high mass groups to the XRB spectral intensity, for 
Galactic latitudes $b>0^{\circ}$ (purple striped curve), 
$b>10^{\circ}$ (black dash-dotted curve), $b>30^{\circ}$ (red dashed 
curve), and the Lockman hole (blue solid curve)
\label{different_absorptions}}
\end{figure}

Results for the average column density at different regions marked by the
lower limit of the Galactic latitude are summarized in table~\ref{column
density}.
\begin{table}
\caption{The average column density at latitudes higher than $b$.
The Lockman hole is situated at $\alpha=10^h45^m$, 
$\delta=57^{\circ}20^{'}$ (Lockman et al. 1986) 
\label{column density}}
\begin{center}
\begin{tabular}{|c|c|}
\hline
   $b$ [$^{\circ}$]              &  $\langle N_H(>b) \rangle$ [$10^{20}$ cm$^{-2}$]\\
\hline
$  0  $      & $13.9\pm0.3$          \\
$  10 $      & $6.05\pm0.06$         \\
$  30 $      & $3.54\pm0.04$         \\
Lockman hole & $0.45\pm0.05$         \\ 
\hline
\end{tabular}
\end{center}
\end{table}
The contribution of clusters to the XRB with $A=0.3$, taking into 
account the effect of Galactic absorption, but excluding the effect of 
mergers, is illustrated in Fig.~\ref{different_absorptions}.

\subsection{IC Gas Density and Temperature Profiles} 
\label{gas profiles}

The density profile of IC gas 
is usually represented by a $\beta$ model, which fits well 
X-ray and SZ measurements (though occasionally a double 
$\beta$ model provides a better fit). However,  
the assumption of isothermality is known to be unrealistic, 
especially in the outer regions of clusters. A polytropic 
equation of state provides a more realistic temperature profile, since 
at $r>>r_c$, $n(r)\propto r^{-3\beta}$ and $T(r)\propto 
r^{-3\beta(\gamma-1)}$. 

We can use insight on the profile of the gas entropy parameter, 
$K\equiv T\, n^{-2/3}$, to select the relevant range of $\gamma$ 
values. In the outer regions of clusters, $r\gtrsim 0.1R_{\rm vir}$, 
beyond the cool core, the entropy is entirely due to gravitational 
processes, and its profile is expected to be a featureless power law 
approaching $K \propto r^{1.1}$ (Tozzi \& Norman 2001; Voit, Kay, \& 
Bryan 2005). However, a 
recent model-independent analysis of X-ray and lensing measurements of 
A1689 yielded $0.82\pm 0.02$ for the power-law index of the profile 
(L08a). For isothermal gas, $\beta=2/3$ implies 
$K\propto r^{4/3}$. This value of the index is higher than the 
two aforementioned values of 1.1 and 0.82. Thus, in addition to not 
being realistic at large radii, the isothermal case is also inconsistent 
with the entropy profile when $\beta=2/3$ is adopted. In order for the 
entropy profile to have an index of 0.82, $\gamma$ has to be 
$\simeq 1.25$, which is in good agreement with both 
ASCA measurements of 30 nearby clusters (Markevitch et al. 1998) 
and numerical simulations (Loken et al. 2002). 

\begin{figure}
\centering
\epsfig{file=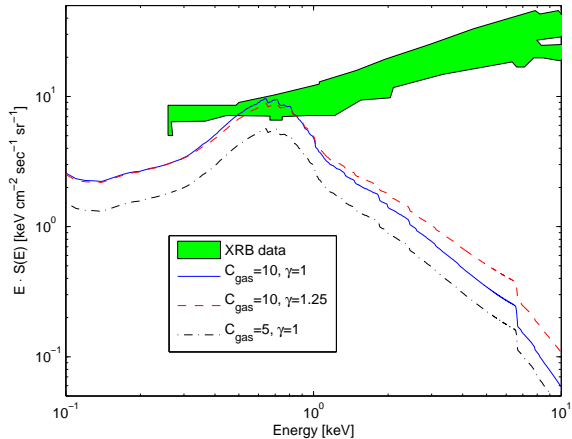, width=8.5cm, clip=}
\caption{Dependence of the cluster CXRB contribution on different values of 
$C_{gas}$ and $\gamma$.
\label{different_gamma_values}}
\end{figure}

In fig.~\ref{different_gamma_values} we explore the cluster contribution 
to the CXRB for different values of $C_{gas}$ and $\gamma$.
For $C_{gas}=10$ and $\gamma=1.25$ the X-ray intensity is higher at 
$E> 1$ keV as compared with its value for $\gamma=1$, reflecting 
the higher central temperature. 
Although the temperature at $r\gtrsim r_{vir}$ is lower than in the 
$\gamma=1$ case, the radial integration (to $r_{vir}$) 
only results in a boost of the intensity level 
at high energies. For $C_{gas}=5$, there 
is no significant difference between the $\gamma=1$ and $\gamma=1.25$ 
cases, a result of which can be explained taking into account the fact 
that the change in the central temperature (which depends on $C_{gas}$) 
is lower, and that the core radius of the gas is bigger so that the range 
of difference between the core and the virial radii is smaller.

\subsection{Impact of cluster mergers}  
\label{Impact of cluster mergers}

The formation of structure in CDM models is hierarchical; clusters evolve 
through mergers of subclumps. During these merger events the temperature 
and luminosity of IC gas are temporarily boosted up with respect to the 
corresponding equilibrium values, in a manner that has been characterized 
by hydrodynamical simulations. Neglecting this boost may lead to an 
underestimation of the cluster contribution to the CXRB. Randall, 
Sarazin, \& Ricker (2002, hereafter RSR02) employed a semianalytic 
model to estimate the effect of merger boosts on the X-ray luminosity 
function (XLF) and temperature function (TF). A quantification of the 
luminosity and temperature spikes in individual mergers was made based on 
N-body/hydrodynamical simulations, whereas the statistics of the merger 
history were determined from Extended PS (EPS) merger trees. 
RSR02 used these merger-boosted TFs and XLFs to infer $\sigma_8$ and 
$\Omega_m$ by converting them to PS mass functions and comparing with 
local ($z=0$) and distant (either $z=0.5$ or $z=1$) cluster samples. 
For the $\Lambda$CDM model an increment of 20\% in $\sigma_8$ was 
deduced by fitting to both the TF and XLF. On the other hand, results 
for $\Omega_m$ were inconclusive; fitting to the TF and XLF exhibited 
a decrement and an increment, respectively. In fact, the results obtained 
from the XLF fit were less reliable, as indicated by the fact that 
$\Omega_m$ inferred from the non-boosted functions was lower by 
60\% than the actual value used to construct the merger trees. 
The reason for the lower degree of reliability associated with the fit 
to the XLF is due to the fact that the range of X-ray luminosities is 
broader than that of X-ray temperatures; put otherwise, the X-ray
luminosity-temperature scaling is steeper than a linear relation. 
As a result, the overall effect of luminosity boosts on the XLF is not as
strong as the effect of temperature boosts on the TF.

\begin{figure} 
\centering
\epsfig{file=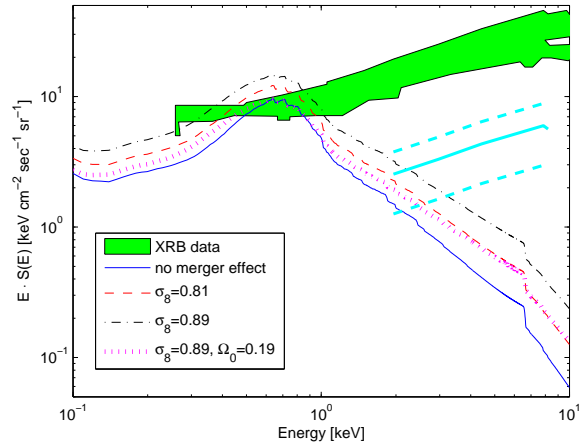, width=8.5cm, clip=}
\caption{The cluster contribution to the  
CXRB including the effect of mergers. Plotted are the 
XRB data (green area), the cluster contribution 
without the effect of mergers 
(blue solid curve), and the corresponding contributions 
when their effect was modelled by selecting values of $\sigma_8$ 
higher by 10\% and 20\% (red dashed and black dot-dashed curves, 
respectively), and by increasing $\sigma_8$ by 20\% but lowering 
$\Omega_m$ by 20\% (purple stripped curve). Also shown is the currently 
estimated upper limit on the residual (i.e., after accounting for 
emission from AGN) 2-8 keV emission and its 1-$\sigma$ range (cyan solid 
and dashed lines, respectively).
\label{merger_effect}}
\end{figure}

The fitted $\sigma_8$ to the boosted TFs was higher by 20\% than the 
value used to construct the trees, whereas with the non-boosted TFs it 
was higher by only 10\% (a difference which is due to the approximation 
used to determine the cluster formation redshift). 
In Fig.~\ref{merger_effect} we compare our calculated cluster contribution 
to the CXRB including the effect of mergers with the data. 
The impact of mergers was modeled in three different ways by selecting 
values of $\sigma_8$ higher by 10\% ($\sigma_8=0.81$) and 20\% 
($\sigma_8=0.89$), and by adopting a combination of $\sigma_8$ higher 
by 20\% ($\sigma_8=0.89$) and of $\Omega_m$ lower by 20\% 
($\Omega_m=0.19$).
As is clear from the figure, the third choice, i.e. the one combining a
higher $\sigma_8$ and a lower $\Omega_m$, would seem more likely, 
simply by virtue of its consistency with the data. In fact, this scenario 
was also preferred by RSR02, who noted that the effect of merger boosts 
on $\sigma_8$ and $\Omega_m$ depends somewhat on the detailed method 
used to determine the TFs and XLFs, and the criteria used to fit them. 
It should be noted that this holds for $C_{gas}=10$; for 
$C_{gas}=5$, none of the models exceed the observed intensity range. 
However, this is no longer the case if $\sigma_8$ is higher, as is alluded 
to in \textsection~\ref{Discussion}. In addition, in the the $2-8$ keV 
energy band high values of $\sigma_8$ boost the cluster CXRB contribution 
above the 1-$\sigma$ upper limit of the residual XRB, which was calculated 
by multiplying the measured XRB upper bound (in the $2-8$ keV band) by the 
residual fraction determined by Bauer et al.\ (2004). 

Results for merger boosts that emerge from various works differ in 
several respects. Bullock et al. (2001) found in their simulations a 
value of $\lambda_{med}$, the spin parameter (Peebles 1971), lower 
than $0.05$, the value used in RSR02. This would slightly increase the 
effect of merger boosts. On the other hand, preheating and 
non-gravitational effects may somewhat reduce the effect of merger 
boosts since the fractional increase in the X-ray luminosity and 
temperature would be smaller if non-gravitational heating is considered. 
This effect would dominate in low-mass clusters (Ponman et al. 1996).

\subsection{Non-thermal pressure} 
\label{Non-thermal pressure}

The hydrostatic equation (HE) for an ideal gas can be integrated from 
a given radius $r$ (assuming spherical symmetry) out to some limiting 
radius 
\begin{equation} \left[ \rho_{gas}(r)T(r)\right] \Big|^{\infty}_{r} =
C \int^{\infty}_{r}\frac{GM(\leq r')\mu m_{p} \rho_{gas}(r')
dr'}{k_{B}r'^{2}},
\label{eq: HEE}
\end{equation}
where $\rho_{gas}$ and $T$ are the gas mass 
density and temperature, respectively, $M( \leq r)$ is the total mass 
within radius $r$. If the only contribution to the pressure - which 
balances gravity - is the thermal gas, $C=1$ (e.g., Binney \& Tremaine 
1987).

In our work so far we have assumed that clusters are in hydrostatic 
equilibrium, and that gas pressure is strictly thermal. 
However, this may not necessarily be the case, at least in some clusters. 
Accordingly, the impact of having a second pressure component was assessed 
heuristically (ignoring other possible ramifications of the 
source of the additional pressure), simply by 
gauging the lower IC gas contribution by taking $C=0.9$ and 
$C=0.8$ (assuming, rather realistically, that a non-thermal - such 
as turbulent and energetic particle - component does not exceed 20\% 
of the total pressure).
\begin{figure} 
\centering
\epsfig{file=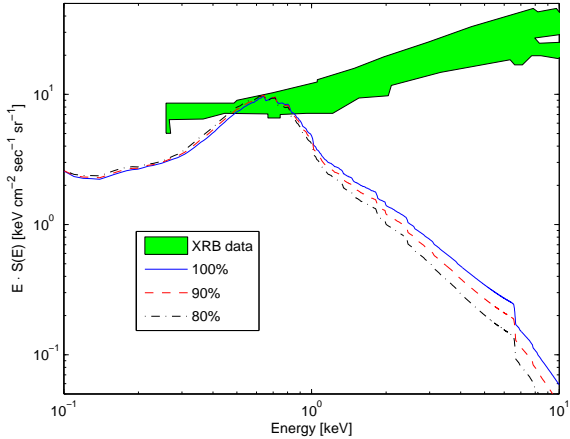, width=8.5cm, clip=}
\caption{The cluster CXRB contribution for a purely thermal, 
and for the case when the thermal pressure is only 90 \%, or 80\% 
(blue solid, red dashed, black dot-dashes curves, respectively).
\label{Other_mechnisms_to_the_pressure}}
\end{figure}   
We can see from Fig.~\ref{Other_mechnisms_to_the_pressure} that 
as $C$ decreases the cluster contribution is lowered at energies 
$>1$ keV, with practically no change at lower energies.

\subsection{Seth \& Tormen mass function}
\label{Modified Press-Schechter mass function}

\begin{figure} 
\centering
\epsfig{file=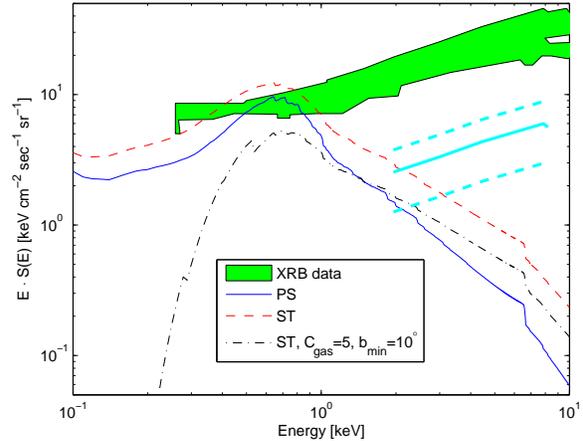, width=8.5cm, clip=}
\caption{The cluster CXRB contribution 
calculated with the PS (solid line), ST (dashed line) mass functions; 
also shown is the result for ST mass function with $C_{gas}=5$ and 
Galactic absorption with $b_{min}=10^{\circ}$ 
(dash-dotted line). Also shown is the currently estimated upper limit on the residual 
(i.e., after accounting for emission from AGN) 2-8 keV emission 
and its 1-$\sigma$ range (cyan solid and dashed lines, respectively).
\label{PS_Vs_ST}}
\end{figure}

The PS mass function, which is derived under the naive assumption of 
spherical collapse, 
underpredicts the abundance of higher mass clusters, and overpredicts 
the abundance of lower mass systems, 
with respect to those observed in N-body simulations. Such simulations 
manifest a clear non-spherical collapse behaviour. Sheth \& Tormen (1999, 
hereafter ST; see also Maggiore \& Riotto 2009) developed a modified mass 
function, which provides a better 
agreement with results of simulations (Sheth, Mo, \& Tormen 2001):
\begin{equation}
F(\nu) = A\sqrt{\frac{2a}{\pi}}\left[ 1+\left(\frac{\nu^2}{a} 
\right)^p\right]e^{-\frac{a\nu^2}{2}}\frac{\nu}{\sigma},
\end{equation} where $A=0.3222$, $a=0.707$, and $p=0.3$.
The PS mass function is reproduced when $A=0.5$, $a=1$, and $p=0$.
In Fig.~\ref{PS_Vs_ST} we plot the CXRB spectrum in the mass range 
$1.1\cdot10^{13}$ h$_{0.7}^{-1}$ $M_{\odot}$ - $7.3\cdot10^{15}$ 
h$_{0.7}^{-1}$ $M_{\odot}$, using the PS (solid line) and ST (dashed 
line) mass functions, and the ST mass function with $C_{gas}=5$ and 
Galactic absorption with $b_{min}=10^{\circ}$ 
(dash-dotted line). 

Since the ST mass function generates a higher population of high-mass 
halos with respect to the PS mass function in almost the entire 
considered mass range, it gives rise to a significant increase of the 
predicted spectrum. If we neglect Galactic absorption, the computed 
levels exceed the measured XRB range; if Galactic absorption is included, 
the power spectrum peaks at approximately the same level as the unabsorbed 
spectrum generated with the PS mass function. Since not all of the XRB in 
the $0.4\lesssim E \lesssim 1$ keV energy range is due to clusters, 
their partial contribution actually puts a very strong constraint on the 
value of $C_{gas}$: As is clear from the figure, the ST mass function 
yields intensity levels which are roughly half of the observational upper 
limit, if Galactic absorption with $b_{min}=10^{\circ}$ 
and $C_{gas}=5$ are adopted. Note that the residual emission in the $2-8$ keV 
band yields an upper limit of $\sim 10$ on $C_{gas}$.

\subsection{Alternative cosmological models} 
\label{Alternative cosmological models}

\begin{figure} 
\centering
\epsfig{file=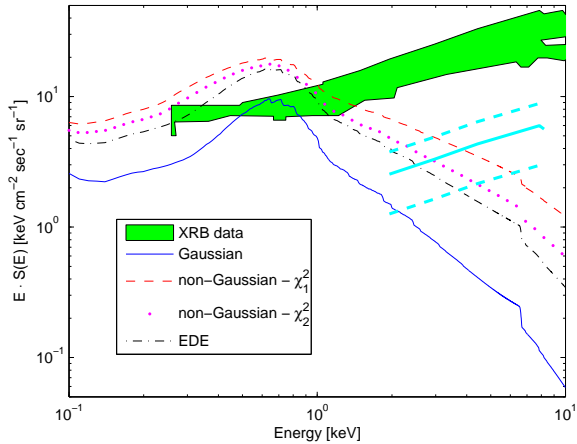, width=8.5cm, clip=}
\caption{X-ray spectral intensity curves produced by a cluster population 
derived from a PS mass function with 
a Gaussian (solid line), $\chi^2_1$ (dashed line), and $\chi^2_2$ 
(dotted line) PDFs, and an EDE model (dot-dashed line). 
Also shown is the currently estimated upper limit on the residual 
(i.e., after accounting for emission from AGN) 2-8 keV emission 
and its 1-$\sigma$ range (cyan solid and dashed lines, respectively).
\label{G_Vs_nonG_n_EDE}}
\end{figure}

As discussed in the \textsection~\ref{Introduction}, there is considerable 
interest in viable alternative cosmological models that predict enhanced 
power on the scale of clusters, and their earlier formation than in 
$\Lambda$CDM. In this section we determine the CXRB cluster  
contribution in the NG and EDE models specified above. In Fig.~ 
\ref{G_Vs_nonG_n_EDE} we compare the X-ray intensity spectra 
produced by a cluster population derived from a PS mass function with 
a Gaussian (solid line), $\chi^2_1$ (dashed line), and $\chi^2_2$ 
(dotted line) PDFs, and an EDE model (dot-dashed line). 
As is clear from Fig.~\ref{G_Vs_nonG_n_EDE}, the predicted intensity 
levels in the NG models for both m=1 and m=2 are 
much higher than the observed range below $\sim 1$ keV. 
Clusters overproduce the XRB 
at a level which is well beyond the range of modeling uncertainties. 
The predicted level of contribution is in fact excessive also in the 
$0.5-2$ keV band, where clusters are presumed to contribute at most 
10-20\% of the measured XRB intensity. In the $2-8$ keV energy band the 
predicted intensity levels in these models are 
higher than the 1-$\sigma$ upper limit of the residual XRB.
This interesting result strengthens the conclusion of 
Sadeh, Rephaeli \& Silk (2007) who showed that the level of CMB anisotropy 
induced by the S-Z effect is significantly higher than the current 
upper limits set by the BIMA experiment. Based on these excess intensity 
and power levels, these particular NG models seem to be ruled out.

\begin{figure} 
\centering
\epsfig{file=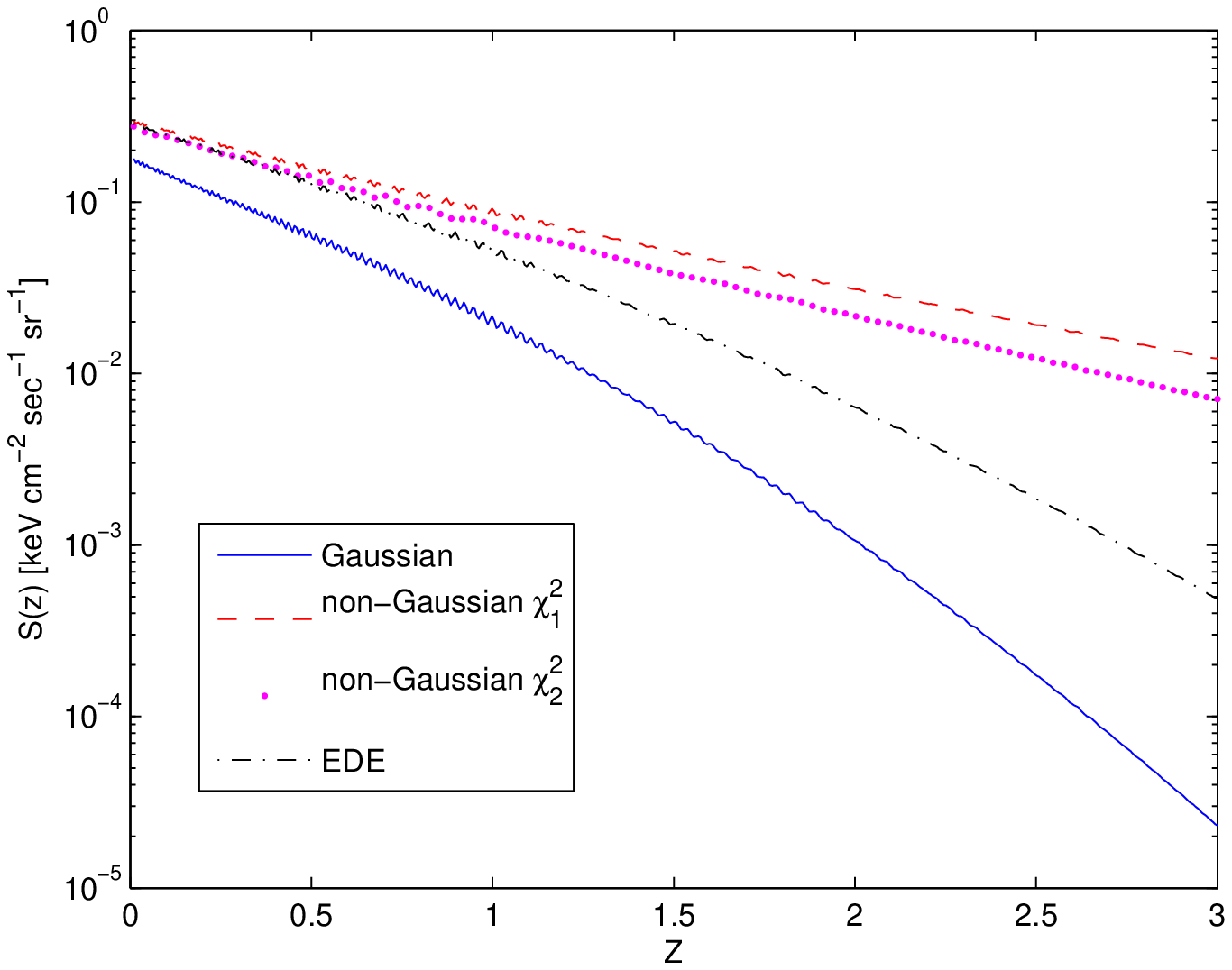, width=8.cm, clip=}
\epsfig{file=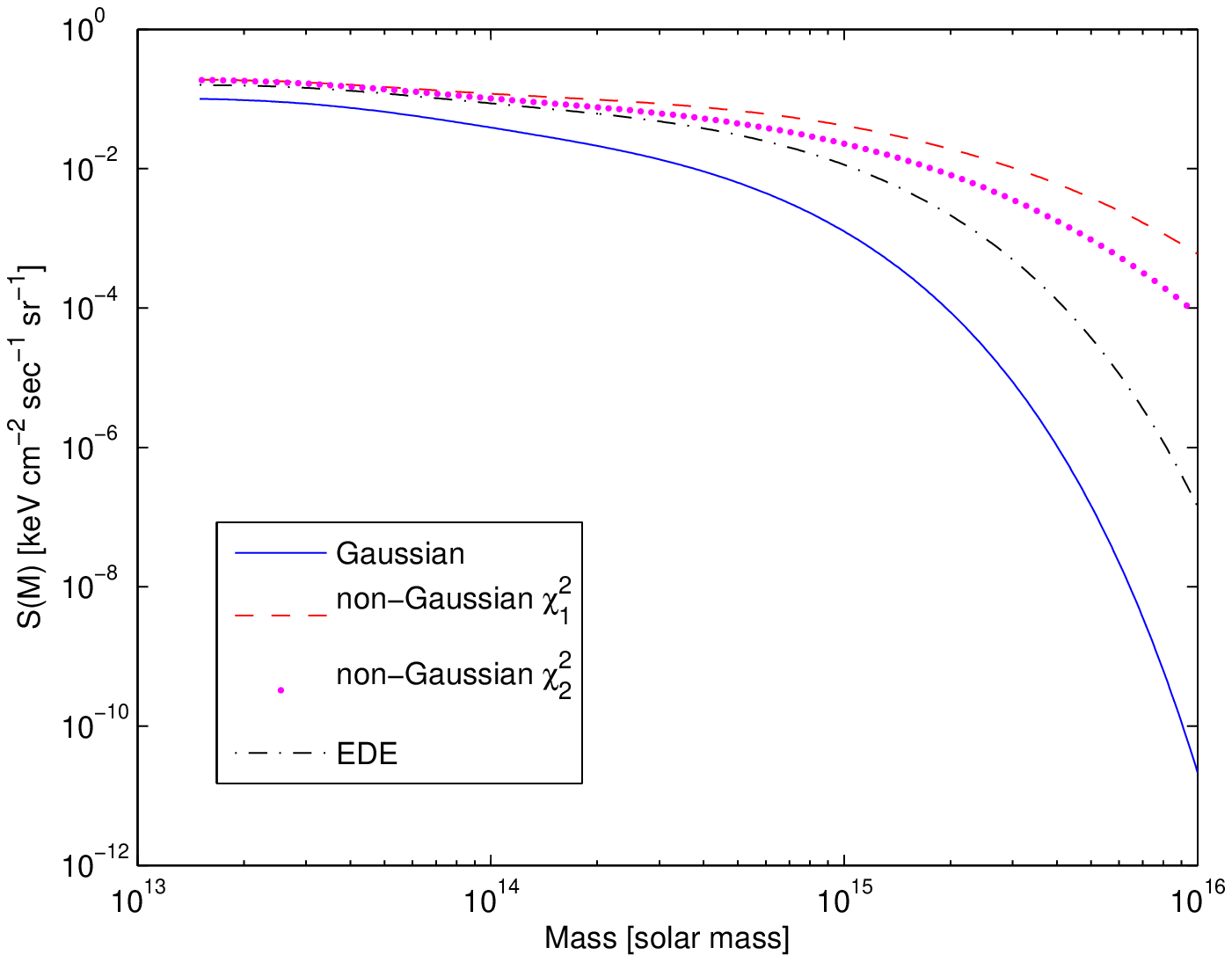, width=8.cm, clip=}
\caption{The redshift (top panel) and mass (bottom panel) distributions 
of the CXRB spectral intensity contributed by clusters are shown for the 
Gaussian (blue solid curve), $\chi^2_1$ (red dashed curve), $\chi^2_2$ 
(purple dotted curve) PDFs, and for the EDE model (black dot-dashed curve).
\label{S_Vs_mass_n_redshift}}
\end{figure}

The degree at which predictions of the EDE model may be reconciled 
with the observational constraints depends on the normalization of the 
matter power spectrum. With the same $\sigma_8$ as derived from WMAP, the 
intensity bump exceeds the XRB data, similar as in the non-Gaussian 
models. However, when we determine the value of $\sigma_8$ by requiring 
that the cumulative halo number at $z=0$ in this model matches the 
corresponding value in the $\Lambda$ CDM we obtain essentially the same 
spectrum. The predictions for the surface brightness distribution as a 
function of mass and redshift are shown in 
Fig.~\ref{S_Vs_mass_n_redshift} for the three models.

The level of tension between predictions of the EDE model and the CXRB 
measurements depends very much on the value of $\sigma_8$. Normalizing $\sigma_8$ 
such that the cumulative halo number at $z=0$ in this model is roughly 
equal to the corresponding value in $\Lambda$CDM leads to similar 
predicted spectra. This 
conclusion is valid even if we adopt $C_{gas}=5$, in addition to using 
the WMAP5-deduced $\sigma_8$, which is higher by $\sim 10$ \% than the 
corresponding WMAP3 value (adopted here). The impact of the higher value 
of $\sigma_8$ is illustrated in Fig.~\ref{merger_effect}.

\section{Discussion}
\label{Discussion}

Significant advancements in sensitivity and spatial resolution have 
resulted in an improved ability to resolve the XRB into its different 
components and sources. The combination of high quality data and 
enhanced-accuracy calculations can be used in order to constrain 
different aspects which affect directly and indirectly the XRB levels. 
As elaborated upon in \textsection~\ref{cluster_CXRB_contribution}, the 
spectral bump in the band $0.4 \lesssim E\lesssim 1$ keV, an 
important characteristic of line emission from high-mass groups and poor 
clusters, constitutes a particularly sensitive probe; models which boost 
the emission in this energy band might exceed the XRB data. Another 
important observational constraint includes the XRB levels in the 
hard energy band, $2-10$ keV. It seems to have been established that 
AGN constitute the major contributors in this energy band. Consequently, 
models in which the cluster contribution is appreciably higher than $\sim 10\%-20\%$ 
of the total CXRB intensity would likely be non-viable. Moreover, the 
combination of extensive measurements of the CXRB and detailed 
predictions of the contributions of AGN, clusters, and starburst galaxies, 
provide a diagnostic tool of the number density and evolution of these 
sources, as well as a probe to set constraints on alternative 
cosmological models. 

We demonstrated in this paper that meaningful constraints can be 
set on alternative cosmological models by contrasting the predicted 
contribution of clusters to the CXRB in non-Gaussian $\chi^2_m$ and EDE 
models with the measurements. As discussed in \textsection~\ref{Introduction}, 
there seems to be some observational evidence for 
earlier cluster formation and higher abundances (as compared with those 
in $\Lambda$CDM) that are predicted in these models. Our basic 
result is that the cluster contribution in the $\chi^2_1$ and 
$\chi^2_2$ models is excessively high at low ($\lesssim 1$ keV) 
energies, so much so that these models could probably be ruled out. 
Moreover, in the high energy band, $2-10$ keV, the predicted spectral 
intensity attains unrealistically high levels by virtue of the fact that 
most of the CXRB at these energies originates in AGN. 
The viability of the EDE model considered here depends very much on 
how exactly this model is normalized with respect to the standard 
model. With the same $\sigma_8$ as derived from WMAP5 values, the 
predicted energy bump in this EDE model exceeds the measured range, 
as is the case in the non-Gaussian models. However, normalizing 
$\sigma_8$ such that the cumulative halo number at $z=0$ in this model 
matches the corresponding value in $\Lambda$CDM, yields nearly the same 
spectrum. These conclusions are still valid even if we adopt 
$C_{gas}=5$ and Galactic absorption with $b_{min}=10^{\circ}$, along with 
ST mass function and the value of $\sigma_8$ deduced from WMAP5 measurements, 
which is higher by $\sim 10$ \% than the corresponding WMAP3 value. The 
effect of this increment in $\sigma_8$ is illustrated in Fig.~\ref{merger_effect}. 

We also showed how the CXRB can be used to constrain merger scenarios. 
We estimated the impact of mergers heuristically by determining the 
resulting boost in the CXRB in terms of the change in $\sigma_8$ and 
$\Omega_m$, following the work of RSR02. CXRB levels calculated for 
two of the three models exceed the observational data in the spectral 
bump region. The $20$\% increment of $\sigma_8$ leads to an 
overproduction of the XRB intensity levels, also when adopting 
$C_{gas}=5$ and the WMAP5 parameters, whereas a $10$\% boost is in 
marginal agreement with the data (for a more detailed account of the 
models see \textsection~\ref{merger_effect}). The impact of Galactic 
absorption, neglected in these calculations, is discussed in 
\textsection~\ref{Galactic absorption}. 

We explored the impact of non-gravitational heating on the CXRB and 
found only a minor change. While the spectral bump is not affected at 
all, higher energy emission from clusters decreases in proportion 
with the amount of non-gravitational heating. Nonetheless, this 
decrement is small, unless non-gravitational heating is considerable. 
Changing the polytropic index, $\gamma$, from $1$ (isothermal) to $1.25$ 
also does not alter the spectrum considerably, especially for low 
$C_{gas}$. With $C_{gas}=10$ and $\gamma=1.25$ the central temperature 
is higher and the temperature at large radii, $r\gtrsim r_{vir}$ is 
lower than with the $\gamma=1$ case. However, since the integration 
of the mass and gas profiles is performed up to the virial radius, we 
only notice an enhancement of the spectrum levels at high energies. 
For $C_{gas}=5$ there is almost no difference between the $\gamma=1$ 
and $\gamma=1.25$ cases.

It was shown that inclusion of all relevant emission processes (i.e., 
not just Bremsstrahlung) is very important. Especially important is 
line emission, which generates the $0.4 \lesssim E\lesssim 1$ keV 
spectral bump. We calculated the spectrum taking mean abundances  
of 0 and 0.3 solar. In clusters the metallicity is typically $\sim 0.2$ 
solar outside the central region (Baldi et al. 2007; Leccardi \& Molendi 
2008). The Fe abundance is nearly similar in all hot clusters, $>4.5$ keV 
(Mushotzky 2004 and reference within), even though medium temperature 
clusters, $2-4$ keV, have abundances of $\sim 0.4$ (Baumgartner et al. 
2005). There seems to be a slight redshift dependence; XMM and Chandra 
show no evolution in Fe abundances up to $z\approx0.8$ (e.g., Mushotzky 
2004 and reference within). However, a certain degree of enrichment must 
have occurred. Balestra et al. (2007) found that the average iron content 
of IC gas at $z=0$ is a factor $\sim 2$ higher than at $z=\sim1.2$. 
Several low-redshift, bright X-ray groups were shown by Finoguenov \& 
Ponman (1999) to have Fe abundances of $\lesssim 0.3$ at radii higher 
than $\sim 100$ kpc. For almost any kind of abundance gradient and 
evolution the CXRB contributions of high mass groups and clusters is 
between our derived values for $A=0$ and $A=0.3$.

The impact of using a mass function different than PS was explored by 
repeating the calculations with the ST mass function, which is 
characterised by higher abundances of high-mass halos in the mass range 
$M>4\cdot10^{13}$ h$_{0.7}^{-1}$ M$_{\odot}$ with respect to the 
corresponding numbers predicted by the PS mass function. The enhanced 
abundances result in appreciably higher intensity levels that possibly 
exceed the measured XRB range for high values of $C_{gas}$ even in 
$\Lambda$CDM. Clearly, given that not all of the XRB in the 
$0.4\lesssim E \lesssim 1$ keV energy range is due to high-mass 
groups and clusters, this implied excess provides a very tight 
constraint on the maximum value of $C_{gas}$. (Note that the mass range 
of the Jenkins et al. (2001) mass function, which was deduced from 
N-body simulations, is irrelevant to our work here.) If $C_{gas}\approx C_{DM}$,
where $C_{DM}$ is the concentration parameter, the XRB can put strong 
constrains on the value of the concentration parameter.

Finally, we note that whereas we have accounted for emission from groups 
and clusters, our treatment does not include emission from the 
filamentary WHIM. With temperatures $\sim 10^6$ or slightly higher, 
WHIM emission could also contribute somewhat to the CXRB. Clearly, any 
additional contribution will only strengthen our conclusions on the 
viability of alternative models in which the predicted spectral bump 
exceeds the observed range.

\section*{ACKNOWLEDGMENT}

We wish to thank the anonymous referee for a thorough reading of  
a previous version of the paper and for several useful suggestions.

\end{document}